# Structure-Dependent Fluorescence Efficiencies of Individual Single-Walled Carbon Nanotubes

*Dmitri A. Tsyboulski[1], John-David R. Rocha[1], Sergei M. Bachilo[1], Laurent Cognet[1,2] and R. Bruce Weisman[1,*]*

[1]Department of Chemistry, Richard E. Smalley Institute for Nanoscale Science and Technology, and Center for Biological and Environmental Nanotechnology, Rice University, 6100 Main Street, Houston, Texas 77005

[2]Centre de Physique Moléculaire Optique et Hertzienne, Université Bordeaux 1, and CNRS, Talence, F-33405 France

AUTHOR EMAIL ADDRESS: weisman@rice.edu

ABSTRACT: Single-nanotube photometry was used to measure the product of absorption cross-section and fluorescence quantum yield for 12 (*n*,*m*) structural species of semiconducting SWNTs in aqueous SDBS suspension. These products ranged from 1.7 to 4.5 × $10^{-19}$ $cm^2$/C atom, generally increasing with optical band gap as described by the energy gap law. The findings suggest fluorescent quantum yields of ~8% for the brightest, (10,2) species and introduce the empirical calibration factors needed to deduce quantitative (*n*,*m*) distributions from bulk fluorimetric intensities.

All current methods for producing single-walled carbon nanotubes (SWNTs) give mixtures of (*n*,*m*) structural types, and therefore a variety of diameters, chiral angles, and electronic characters. Since the discovery in 2002 of near-IR fluorescence (photoluminescence) from semiconducting SWNTs and the subsequent assignment of well defined spectral peaks to specific (*n*,*m*) species,[1-3] fluorimetry has emerged as the most promising approach for analyzing nanotube bulk mixtures because of its high sensitivity and its ability to detect many species using only a few excitation wavelengths. Fluorimetry readily provides qualitative analysis, identifying semiconducting (*n*,*m*) species through their securely assigned absorption and emission wavelengths. However, quantitative analysis to reveal their abundances requires knowledge of the structure-dependent factors that connect observed fluorimetric intensities to the actual species concentrations. These factors are also fundamental to understanding nanotube photophysics. We report here the use of single-nanotube microphotometry for the first experimental measurement of fluorescence action cross-sections of 12 semiconducting species. Our results provide an experimental foundation for improved theoretical models of nanotube photophysics and represent a significant step towards the goal of structure-resolved quantitative analysis of bulk SWNT samples.

Fluorescence intensity is proportional to the concentration of a species, its absorption cross-section (or absorptivity) at the excitation wavelength, and its fluorescence quantum yield. The term fluorescence action cross-section describes the product of fluorescence quantum yield and absorption cross-section (or absorptivity) normalized for the number of carbon atoms. The simplest approach to finding this action cross-section as a function of structure would be to make careful fluorimetric measurements on reference samples of pure or well known (*n,m*) compositions. Unfortunately, such reference samples are not yet available. However, single-molecule optical methods allow the study of individual, (*n,m*)-selected nanotubes within an inhomogeneous mixture to reveal the photophysical properties of specific species.[4-9] We therefore used single-nanotube near-IR microscopy to measure the absolute fluorescence action cross-sections of a set of semiconducting SWNTs.

The factors influencing action cross-sections can be divided into two types. The first type is nanotube structure (diameter and chiral angle, or equivalently (*n,m*)), which controls the intrinsic electronic and vibrational properties of pristine nanotubes and thereby leads to structure-dependent variations in fluorescence efficiency. Nanotube structure should solely determine the action cross-section of disaggregated, perfect, and infinitely long SWNTs in a uniform, well defined environment. Theoretical methods should be able to model and predict these intrinsic variations. The second type of factor includes all non-ideal influences such as finite length; growth defects; aggregation and other environmental irregularities; and physical or chemical damage caused by sample processing. We presume that these extrinsic factors tend to quench fluorescence, thereby lowering emissive quantum yields and action cross-sections. Although both intrinsic and extrinsic factors will affect real samples, we have designed our experiments to isolate and measure the intrinsic action cross-sections. Measurements were made in aqueous SDBS suspensions on individual nanotubes that were selected for minimal apparent influence from extrinsic effects.

The spectrally and spatially integrated emission signal $S_{FL}$ detected from a single nanotube excited at wavelength $\lambda_{exc}$ and emitting near peak wavelength $\lambda_{11}$ can be written as:

$$S_{Fl} = \eta(\lambda_{11}) \cdot \alpha(\lambda_{11}) \cdot F(\lambda_{exc}) \cdot \sigma(\lambda_{exc}) \cdot \Phi_{Fl}, \qquad (1)$$

where $\eta(\lambda_{11})$ is the detector sensitivity factor, $\alpha(\lambda_{11})$ is the optical collection efficiency, $F(\lambda_{exc})$ is the excitation photon flux at the sample, and $\sigma(\lambda_{exc})$ is the SWNT absorption cross-section at the excitation wavelength. By measuring $S_{Fl}$, $\eta(\lambda_{11})$, $\alpha(\lambda_{11})$ and $F(\lambda_{exc})$, one can directly access the product of the nanotube absorption cross-section and its fluorescence quantum yield. To determine this action cross-section when the excitation wavelength does not match the peak wavelength of $E_{22}$ absorption, we apply a correction factor of $\sigma(\lambda_{22}^{peak})/\sigma(\lambda_{exc})$. It is expected that fluorescence signals $S_{Fl}$ for a given (*n,m*) species are proportional to length for SWNTs long enough to neglect the quenching of excitons created within ~50 nm (half of their excursion range)[9] of the nanotube ends. End quenching effects in our experiments were minimized by restricting measurements to nanotubes longer than 3 μm.

Samples were prepared by dispersing a few micrograms of raw HiPco material in 5 mL of 1% aqueous SDBS.[5] The sample was ultrasonically treated for 5 s at 7 W using a Microson XL-2000 ultrasonic liquid processor with a 3 mm diameter tip. The pH was adjusted to 8 by adding small amounts of NaOH. Then a 1.5 μL portion of the SWNT suspension was spread between a glass microscope slide and a cover slip. We performed near-IR fluorescence imaging and spectroscopy using an apparatus described previously.[5] The Nikon TE-2000U inverted microscope was equipped with a Nikon PlanApo VC 60×/ 1.4 NA oil-immersion objective. A dichroic beamsplitter and a 946 nm long-pass filter selected emission in the desired near-IR wavelength range. An InGaAs near-IR imager (OMA-V 2D, Roper Scientific) was installed on one output port of the microscope. At another port, the image plane was coupled via a fiber optic cable (100 μm core diameter) to the entrance slit of a J-Y C140 spectrograph with a 512 element InGaAs detector array (OMA-V, Roper Scientific). Emission spectra between 950 and 1580 nm were thereby acquired from selected spatial regions measuring ~1.5 × 1.5 μm at the sample at 60× magnification.

We excited samples with circularly polarized beams from diode lasers emitting at 658, 730 or 785 nm. The circular polarization ellipticity was ≥0.95. A set of neutral density filters was used to attenuate excitation intensities to the range of 90 to 200 W/cm$^2$. We confirmed that the SWNT emission intensity was proportional to excitation intensity within this range and that nonlinear effects were negligible. Excitation intensity was determined with an accuracy of 8% in each measurement run (see Supporting Information). For each excitation laser, four (*n*,*m*) species with near-resonant $E_{22}$ absorptions were chosen for study: (8,3), (7,5), (7,6), (9,5) for 658 nm excitation; (10,2), (9,4), (8,6), (8,7) for 730 nm excitation; (12,1), (11,3), (10,5), (9,7) for 785 nm excitation. We restricted our observations to these near-resonant species so that the off-resonance excitation correction factors $\sigma(\lambda_{22}^{peak})/\sigma(\lambda_{exc})$ would fall below 2 and could therefore be estimated very accurately from $E_{22}$ transition widths measured for each species and the wavelength difference between the laser and the $E_{22}$ peak. The $E_{22}$ transitions were found to have Lorentzian shapes and half-widths at half-maximum typically exceeding 25 nm. These widths were nearly insensitive to environment and are assumed to reflect homogeneous broadening in individual nanotubes.[2,4,10,11]

A small fraction of SWNTs in our samples had apparent lengths greater than 3 μm. Figure 1 shows four images of such long nanotubes and the corresponding emission spectra. Following the assignment by Bachilo et al.,[2] the SWNT species in Fig. 1 are identified as (12,1), (11,3), (10,5) and (9,7), having emission maxima at 1170, 1196, 1251, and 1323 nm, respectively. Within our diffraction-limited spatial resolution of ~700 nm, the (12,1), (11,3) and (10,5) nanotubes show nearly uniform brightness along their lengths with only a few dim spots. We will refer to these dim areas as "defects" without speculating as to their nature. By contrast, the (9,7) SWNT in Fig. 1 clearly has many defects. Dim and bright areas within one nanotube typically show identical emission spectra. Based on numerous observations, we classify defect regions as those in which the local intensity is between 0 and 90% of the intensity from the brightest portions of the same nanotube. The measurements described below were made only on individual nanotubes selected for showing few or no defect regions.

For each such SWNT, we determined its peak emission wavelength and spectral width using Lorentzian curve fitting. Full-widths at half-maximum of individual nanotubes were found to fall between 110 and 160 $cm^{-1}$, depending on SWNT diameter, in agreement with a previous report.[12] Bulk samples typically show emission line widths ~30 $cm^{-1}$ larger. For our measurements on individual SWNTs of a given (*n,m*) species, we found that although the emission spectra all had similar line widths, they divided into two groups on the basis of peak position (see Supporting Information). The first group had peaks within 20 $cm^{-1}$ of the average value for that species as determined from bulk measurements. We identify these as isolated SWNTs whose transition frequencies vary within a narrow range because of minor inhomogeneities in surfactant environment. The second group had emission peaks red-shifted by ~50 $cm^{-1}$ or more from the nominal bulk value. We suspect that are nanotubes in small bundles, which would provide a highly polarizable environment and correspondingly reduced transition frequency.[13] In order to restrict our study to individualized SWNTs (the first group), we therefore excluded all SWNTs with spectral positions outside a ±20 cm$^{-1}$ window centered on the bulk emission frequency.

We measured fluorescence intensities from 117 SWNTs that satisfied all of the selection criteria: (1) a near-resonant $E_{22}$ transition; (2) a length of at least 3 μm; (3) a low defect density; (4) a standard $E_{11}$ peak position; and (5) free motion with no attachment to cell walls. The fraction of nanotubes meeting the defect density criterion seemed to vary with the growth batch of SWNTs that was examined. We also found that long nanotubes were relatively scarce among small diameter species, possibly because of more rapid cutting during sonication. For each of the 12 (*n*,*m*) species studied here, measurements were performed on between 5 and 20 different nanotubes. Although many of these did not show perfectly uniform brightness along their lengths, they all had a uniformly emissive segment at least 1.3 μm long, representing 4 pixels at 90× magnification. Because these segments are longer than the 90 nm mean exciton excursion range,[9] we assume that they are unaffected by surrounding defects and thus reveal the intrinsic fluorescence properties of the nanotube. Fluorescence intensities were normalized by measuring brightness per unit length and dividing by the known number of carbon atoms in that length of the particular (*n*,*m*) species. Sets of measurements on different nanotubes of the same (*n*,*m*) species showed standard deviations of ~20% (see Supporting Information). Table 1 summarizes the relevant experimental parameters and average action cross-sections measured for the studied (*n*,*m*) species. We find that values of $\sigma(\lambda_{22}) \cdot \Phi_{Fl}$ for circularly polarized excitation light propagating perpendicularly to the nanotube axis vary with SWNT species from 1.0 to 2.7 × 10$^5$ cm$^2$/mol C (1.7 to 4.5 × 10$^{-19}$ cm$^2$/C atom). Note that these values should be scaled appropriately for absolute measurements in other experimental configurations, including bulk fluid samples in which the nanotubes assume random orientations. As can be seen from Table 1, (10,2) shows the highest action cross-section among the SWNT species studied to date.

Our findings allow the fluorescence quantum yield, $\Phi_{Fl}$, to be deduced if a reliable estimate of $\sigma(\lambda_{22})$ is available. Unfortunately, absolute absorption cross-sections of SWNTs are difficult to measure because most samples contain many (*n,m*) species with large uncertainties in individual or even total SWNT concentrations. Islam et al.[14] reported a polarized absorption peak cross-section of $1.0 \times 10^6$ cm$^2$/mol C for a stretch-aligned bulk sample of SWNTs having an average diameter near 1.3 nm. However, we note that their broad, structureless $E_{22}$ band with spectral width of ~3200 cm$^{-1}$ contained many unresolved absorption peaks from different (*n,m*) species. Our analysis of fluorescence excitation spectra indicates that the $E_{22}$ line width of each nanotube species in this diameter range is $300 \pm 50$ cm$^{-1}$. We therefore expect the peak absorption cross-section of a single 1.3 nm diameter nanotube to exceed the reported value by a factor of ~10, giving $1 \times 10^7$ cm$^2$/mol C. However, because (10,2) nanotubes have a larger $E_{22}$ width of ~450 cm$^{-1}$,[2] their estimated peak polarized absorption cross-section would be reduced to $7 \times 10^6$ cm$^2$/mol C, or $3.5 \times 10^6$ cm$^2$/mol C for unpolarized light. Using this absorption cross-section value and our measured action cross-section, we estimate $\Phi_{Fl}$ for the (10,2) species as ~8%. This value is comparable to the 7% quantum yield recently reported by Lefebvre et al. for individual air-suspended, larger diameter SWNTs.[7] We note that their quantum yield was deduced using the unadjusted absorption cross-section reported by Islam et al.[14] and might therefore have been overestimated. Our estimate may be consistent with the 1% quantum yield measured by Hertel and co-workers[15] in a (6,5)-enriched bulk sample that is probably dominated, like our bulk samples, by short and imperfect nanotubes. An important future goal will be to use new methods to measure reliable, (*n*,*m*)-dependent absorption cross-sections.

Fluorescence action cross-section reflects the product of three photophysical factors: $E_{22}$ peak absorptivity; efficiency of nonradiative decay from the second ($E_{22}$) to the lowest ($E_{11}$) excitonic manifold; and the ratio of radiative to total (radiative plus nonradiative) decay rates from $E_{11}$ to the electronic ground state. Using molecular photophysics as a guide, we expect that the second of these factors will be close to one and the third will show the strongest dependence on nanotube structure through variations in nonradiative decay from $E_{11}$. This relaxation process entails conversion of the $E_{11}$ electronic excitation into multiple phonons in a single step. In molecules, rate constants $k_{nr}$ for the analogous process (internal conversion) within a set of similar compounds have been observed to follow the energy gap law:[16-18]

$$k_{nr} \propto \exp\left(-\nu/\nu_a\right). \qquad (2)$$

Here ν is the optical emission frequency (reflecting the amount of electronic energy dissipated into vibrational energy) and $\nu_a$ is the frequency of a vibrational "accepting" mode, often one of the highest in the molecule. Figure 2 shows our measured action cross-sections (in molar absorptivity units) plotted against emission frequency. Excluding the (7,6), (7,5), and (8,3) species, the frequency variation of action cross-section appears rather smooth and well represented by the solid curve, which is based on eq. 2. In this model fit, the parameter $\nu_a$ is approximately 1970 cm$^{-1}$, a value similar to the ~1600 cm$^{-1}$ G-band that is the highest frequency SWNT fundamental vibrational mode. The following numerical expression for the solid curve represents the variation of fluorescence action cross-section (converted to absorptivity) as a function of emission wavenumber $\bar{\nu}$ (given by $E_{11}/hc$):

$$\varepsilon(\lambda_{22})\Phi_{Fl} = \left(0.93 \text{ M}^{-1}\text{cm}^{-1}\right) \exp\left(\frac{\bar{\nu}}{1970 \text{ cm}^{-1}}\right). \qquad (3)$$

This expression seems especially useful for nanotubes having emission frequencies below 9500 cm$^{-1}$, while other relaxation processes may become increasingly important for smaller diameter species such as (7,5), (7,6), and (8,3) and cause deviations of those points from the model curve. Until those processes can be accurately modeled, we recommend applying specific empirical action cross-section factors from Table 1 rather than values from eq. 3 when analyzing smaller diameter nanotubes listed in the table. We also note that zigzag species have not yet been investigated and may not be accurately described by eq. 3.

To deduce relative concentrations of (*n*,*m*) species from conventional bulk excitation-emission fluorimetric scans on aqueous SDBS nanotube suspensions, the spectrally integrated $E_{11}$ signal (corrected for excitation photon flux and expressed in quantal rather than power units) measured for the $E_{22}$ peak of each species should be divided by the corresponding $\sigma(\lambda_{22})\cdot\Phi_{Fl}$ value listed in Table 1 or deduced from eq. 3. This will give the distribution of carbon atoms in the sample among semiconducting (*n*,*m*) species. In earlier fluorimetric analyses of SWNT samples, it was generally assumed that $\sigma(\lambda_{22})\cdot\Phi_{Fl}$ is independent of (*n*,*m*) over the range of species present in the sample.[19] Applying the empirical $\sigma(\lambda_{22})\cdot\Phi_{Fl}$ factors tends to reduce the relative abundances of smaller diameter SWNTs, leading to an increase in the deduced average nanotube diameter. Figure 3 illustrates the relative abundances of the 12 studied (*n*,*m*) species in a HiPco sample, as fluorimetrically determined with and without the empirical $\sigma(\lambda_{22})\cdot\Phi_{Fl}$ factors.

(The plots are scaled for equal (7,6) abundances.) The average SWNT diameter evaluated from the set of 12 species increases by ~2% as a result of the applying the $\sigma(\lambda_{22}) \cdot \Phi_{Fl}$ factors. It can also be seen that the predominance of near-armchair chiralities is even greater after correction, in agreement with the conclusion reached by Jorio et al. using a combined Raman and fluorescence bulk analysis.[20]

The results represented by Fig. 2 and eq. 3 can be compared to theoretical computations of intrinsic SWNT optical properties. Reich et al.[21] have focused on differences in peak $\sigma(\lambda_{22})$ values that they predict to arise from the broadening of this transition when the $E_{22}$ exciton energy exceeds twice the $E_{11}$ energy. This spectral broadening is expected to appear only for mod(*n-m*,3)=1, or "mod 1," nanotubes and would reduce the brightness of mod 1 species with small chiral angles relative to near-armchair species. These authors conclude that the strong chiral angle dependence of experimental fluorescence intensities reflects this effect rather the underlying abundance distribution. Detailed comparison of the theoretical model with our experimental findings is limited because the current data set includes just three mod 1 SWNTs: (8,7), (9,5), and (7,6). We find them to have very similar fluorimetric brightness values, although the range of their chiral angles (27.8° to 20.6°) may be too small to reveal the predicted effect. Our data also show no significant difference in brightness between (10,5) and (9,5), two species from opposite mod groups but quite similar diameter and chiral angle. Thus, although the predicted systematic brightness difference between mod 1 and mod 2 species of small chiral angles might be confirmed by future experiments, the current results instead suggest that the dominant parameter correlated with fluorimetric brightness is emission frequency. It also appears that the underlying chirality distribution in many samples favors near-armchair species. Structural variations in radiationless decay rates seem not to have been included in the model of Reich et al.

Oyama et al. have used a different theoretical approach to estimate SWNT fluorescence intensities.[22] Their model combines factors for absorption, nonradiative relaxation, and emission processes. However, these workers apparently did not consider variations in $E_{22}$ linewidths and modeled nonradiative relaxation only from $E_{22}$ to $E_{11}$, neglecting the $E_{11}$ to ground state nonradiative decay that seems much more likely to influence fluorescence quantum yields. Their computed (*n*,*m*) dependence of action cross-sections mimics the general trend seen in Fig. 2 but predicts a range of values three times wider than is seen experimentally. Interestingly, the computation finds (10,2) to be the brightest species, in agreement with our experiment. We hope that our empirical results can serve as a benchmark to help refine theoretical models to the level that they can reliably predict action cross-sections for a broad range of (*n*,*m*) species and gain insights into fundamental photophysical processes.

It is clear that the selected SWNT segments display substantially brighter photoluminescence than has been observed in bulk samples. We attribute this difference in luminescence efficiency to extrinsic quenching, in which exciton lifetimes are shortened by nonradiative decay processes absent in perfect nanotubes. This idea is supported by the data of Fig. 4, which compares measurements from adjacent segments within a single SWNT. Although one of these segments is brighter than the other by a factor of ~2, their emission spectra are identical (Fig. 4b). When emission intensities from the segments are measured as a function of excitation intensity, different behaviors are found. At the lowest excitation levels both show emission intensities that are linearly proportional to excitation, although with different slopes. Fig 4c shows emission vs. excitation data that have been scaled to match at low intensities. It can be seen that emission from the brighter region of the nanotube shows more pronounced sublinear behavior. This effect was observed in more than 10 different nanotubes having nonuniform emission profiles. The sublinear behavior is known to arise from exciton-exciton annihilation (Auger decay) within the nanotube at increased excitation levels.[23,24] Excitons in the dimmer segment show less evidence of this second-order process because their first-order lifetimes are already reduced by quenching defects.

Nonlinear emission effects are not important in normal SWNT fluorimetric analyses, which involve continuous, low intensity excitation. However, it is clear that fluorescence signals from bulk samples can be substantially reduced by extrinsic quenching effects that lead to dim or dark nanotube segments. Do such extrinsic effects change the apparent distribution of relative ($n,m$) concentrations? If the effects act uniformly on the various species, then a sample's overall fluorescence emission spectrum will simply be reduced by a constant factor, with no change in the deduced species distribution. However, one can imagine extrinsic quenching that varies with ($n,m$). For example, if ultrasonic processing leads to average SWNT lengths that vary with diameter, then end quenching effects will make the shorter species relatively dimmer and their concentrations deduced from fluorimetry will be underestimated. Further research is needed to explore such structure dependence of extrinsic quenching in nanotubes and devise ways to minimize or compensate for them. In the meantime, we suggest that the empirical intrinsic action cross-sections described in Table 1 and eq. 3 should be applied as provisional calibration factors to provide the best currently available estimates of (*n,m*) distributions based on bulk fluorimetric intensities.

**Supporting Information Available**: Description of instrumental calibration, distributions of spectral peak positions and widths, and corrections for nonuniform emission profiles of individual nanotubes. This material is available free of charge via the Internet at http://pubs.acs.org.

ACKNOWLEDGEMENTS This work was supported by grants from the Welch Foundation (C-0807 and L-C-0004), the National Science Foundation (CHE-0314270), the NSF Center for Biological and Environmental Nanotechnology (EEC-0647452), NASA (JSC-NNJ06HC25G), and Applied NanoFluorescence, LLC. The authors are also grateful to J. T. Willerson and J. L. Conyers (Univ. of Texas Health Science Center, Houston) for instrumentation support. L.C. thanks the Fulbright Foundation and the DGA (ERE060016) for financial support. J.-D.R.R. thanks the Rice-Houston AGEP program (NSF Cooperative HRD-0450363) for partial support.

## References

(1) O'Connell, M.; Bachilo, S. M.; Huffman, C. B.; Moore, V.; Strano, M. S.; Haroz, E.; Rialon, K.; Boul, P. J.; Noon, W. H.; Kittrell, C.; Ma, J.; Hauge, R. H.; Weisman, R. B.; Smalley, R. E. *Science* **2002,** *297*, 593-596.

(2) Bachilo, S. M.; Strano, M. S.; Kittrell, C.; Hauge, R. H.; Smalley, R. E.; Weisman, R. B. *Science* **2002,** *298*, 2361-2366.

(3) Weisman, R. B.; Bachilo, S. M. *Nano Lett.* **2003,** *3*, 1235-1238.

(4) Lefebvre, J.; Fraser, J. M.; Finnie, P.; Homma, Y. *Phys. Rev. B* **2004,** *69*, 075403-1-075403/5.

(5) Tsyboulski, D. A.; Bachilo, S. M.; Weisman, R. B. *Nano Lett.* **2005,** *5*, 975-979.

(6) Hartschuh, A.; Pedrosa, H. N.; Peterson, J.; Huang, L.; Anger, P.; Qian, H.; Meixner, A.; Steinert, M.; Novotny, L.; Krauss, T. D. *ChemPhysChem* **2005,** *6*, 577-582.

(7) Lefebvre, J.; Austing, D. G.; Bond, J.; Finnie, P. *Nano Lett.* **2006,** *6*, 1603-1608.

(8) Berciaud, S.; Cognet, L.; Poulin, P.; Weisman, R. B.; Lounis, B. *Nano Lett* **2007,** *7*, 1203-1207.

(9) Cognet, L.; Tsyboulski, D.; Rocha, J.-D. R.; Doyle, C. D.; Tour, J. M.; Weisman, R. B. *Science* **2007,** *316*, 1465-1468.

(10) Chou, S. G.; Ribiero, H. B.; Barros, E. B.; Santos, A. P.; Nezich, D.; Samsonidze, G. G.; Fantini, C.; Pimenta, M. A.; Jorio, A.; Filho, F. P.; Dresselhaus, M. S.;
16

Dresselhaus, G.; Saito, R.; Zheng, M.; Onoa, G. B.; Semke, E. D.; Swan, A. K.; Unlu, M. S.; Goldberg, B. B. *Chem. Phys. Lett.* **2004,** *397*, 296-301.

(11) Lefebvre, J.; Fraser, J. M.; Homma, Y.; Finnie, P. *Appl. Phys. A* **2004,** *78*, 1107-1110.

(12) Inoue, T.; Matsuda, K.; Murakami, Y.; Maruyama, S.; Kanemitsu, Y. *Phys. Rev. B* **2006,** *73*, 233401-1-233401-4.

(13) Wang, F.; Sfeir, M. Y.; Huang, L.; Huang, X. M. H.; Wu, Y.; Kim, J.; Hone, J.; O'Brien, S.; Brus, L. E.; Heinz, T. F. *Phys. Rev. Lett.* **2006,** *96*, 167401-1-167401-4.

(14) Islam, M. F.; Milkie, D. E.; Kane, C. L.; Yodh, A. G.; Kikkawa, J. M. *Phys. Rev. Lett.* **2004,** *93*, 037404-1-037404-4.

(15) Crochet, J.; Clemens, M.; Hertel, T. *J. Am. Chem. Soc.* **2007,** *129*, 8058-8059.

(16) Englman, R.; Jortner, J. *Molec. Phys.* **1970,** *18*, 145-164.

(17) Bachilo, S. M.; Spangler, C. W.; Gillbro, T. *Chem. Phys. Lett.* **1998,** *283*, 235-242.

(18) Yoshizawa, M.; Aoki, H.; Ue, M.; Hashimoto, H. *Phys. Rev. B* **2003,** *67*, 174302-1-174302-8.

(19) Bachilo, S. M.; Balzano, L.; Herrera, J. E.; Pompeo, F.; Resasco, D. E.; Weisman, R. B. *J. Am. Chem. Soc.* **2003,** *125*, 11186-11187.

(20) Jorio, A.; Fantini, C.; Pimenta, M. A.; Heller, D. A.; Strano, M. S.; Dresselhaus, M. S.; Oyama, Y.; Jiang, J.; Saito, R. *Appl. Phys. Lett.* **2006,** *88*, 023109-1-023109/3.

(21) Reich, S.; Thomsen, C.; Robertson, J. *Phys. Rev. Lett.* **2005,** *95*, 077402-1-077402/4.

(22) Oyama, Y.; Saito, R.; Sato, K.; Jiang, J.; Samsonidze, G. G.; Gruneis, A.; Miyauchi, Y.; Maryama, S.; Jorio, A.; Dresselhaus, G.; Dresselhaus, M. S. *Carbon* **2006,** *44*, 873-879.

(23) Wang, F.; Dukovic, G.; Knoesel, E.; Brus, L. E.; Heinz, T. F. *Phys. Rev. B* **2004,** *70*, 1-4.

(24) Ma, Y.-Z.; Valkunas, L.; Dexheimer, S. L.; Bachilo, S. M.; Fleming, G. R. *Phys. Rev. Lett.* **2005,** *94*, 157402-1-157402/4.

Table 1. Experimental parameters and measurement results for SWNT fluorescence action cross-sections in circularly excitation light. The last three columns show action cross-sections expressed for absorption units of cross-section per mole of carbon, cross-section per carbon atom, and (base 10) molar absorptivity per mole of carbon.

| $(n,m)$ | $\lambda_{exc}$ (nm) | $\lambda_{22}$ (nm) | $\lambda_{11}$ (nm) | $\sigma(\lambda_{22}^{peak})/\sigma(\lambda_{exc})$ | $\sigma(\lambda_{22}) \cdot \Phi_{Fl}$ ($10^5$ cm$^2$/mol C) | $\sigma(\lambda_{22}) \cdot \Phi_{Fl}$ ($10^{-19}$ cm$^2$/C) | $\varepsilon(\lambda_{22}) \cdot \Phi_{Fl}$ (L mol$_C^{-1}$ cm$^{-1}$) |
|---|---|---|---|---|---|---|---|
| (8,3)  | 658 | 663 | 952  | 1.12 | 2.1 | 3.5 | 91 |
| (7,5)  |     | 644 | 1023 | 1.92 | 1.9 | 3.2 | 83 |
| (7,6)  |     | 647 | 1122 | 1.61 | 1.4 | 2.3 | 61 |
| (9,5)  |     | 671 | 1244 | 1.82 | 1.1 | 1.8 | 48 |
| (10,2) | 730 | 734 | 1053 | 1.09 | 2.7 | 4.5 | 117 |
| (9,4)  |     | 720 | 1101 | 1.49 | 1.9 | 3.2 | 83 |
| (8,6)  |     | 716 | 1172 | 1.89 | 1.7 | 2.8 | 74 |
| (8,7)  |     | 728 | 1267 | 1.02 | 1.1 | 1.8 | 48 |
| (12,1) | 785 | 797 | 1171 | 1.75 | 1.7 | 2.8 | 74 |
| (11,3) |     | 792 | 1197 | 1.27 | 1.8 | 3.0 | 78 |
| (10,5) |     | 786 | 1250 | 1.01 | 1.2 | 2.0 | 52 |
| (9,7)  |     | 790 | 1323 | 1.14 | 1.0 | 1.7 | 43 |

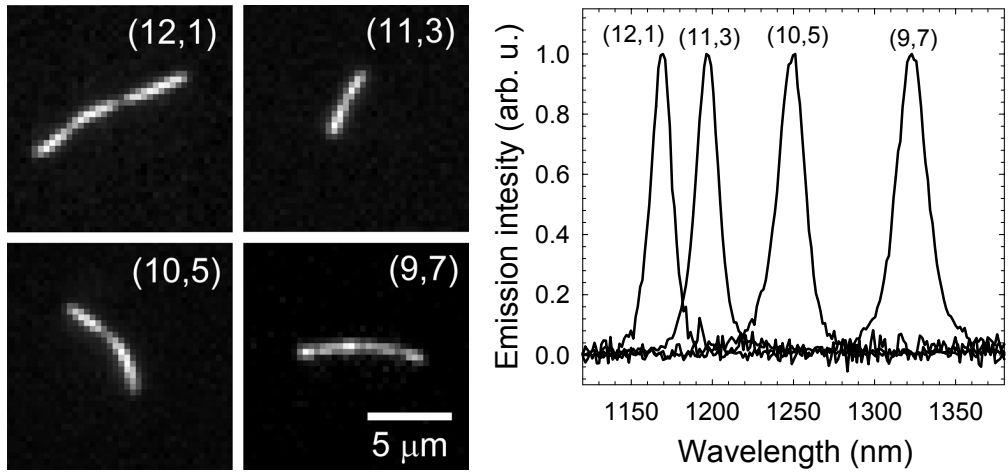

Figure 1. Fluorescence images of SWNTs in aqueous suspension with spatially resolved lengths (left) and the corresponding emission spectra (right). Images were recorded with 50 ms exposures at 90× total magnification and 180 W/cm$^2$ excitation intensity. Spectra were integrated for 5 s.

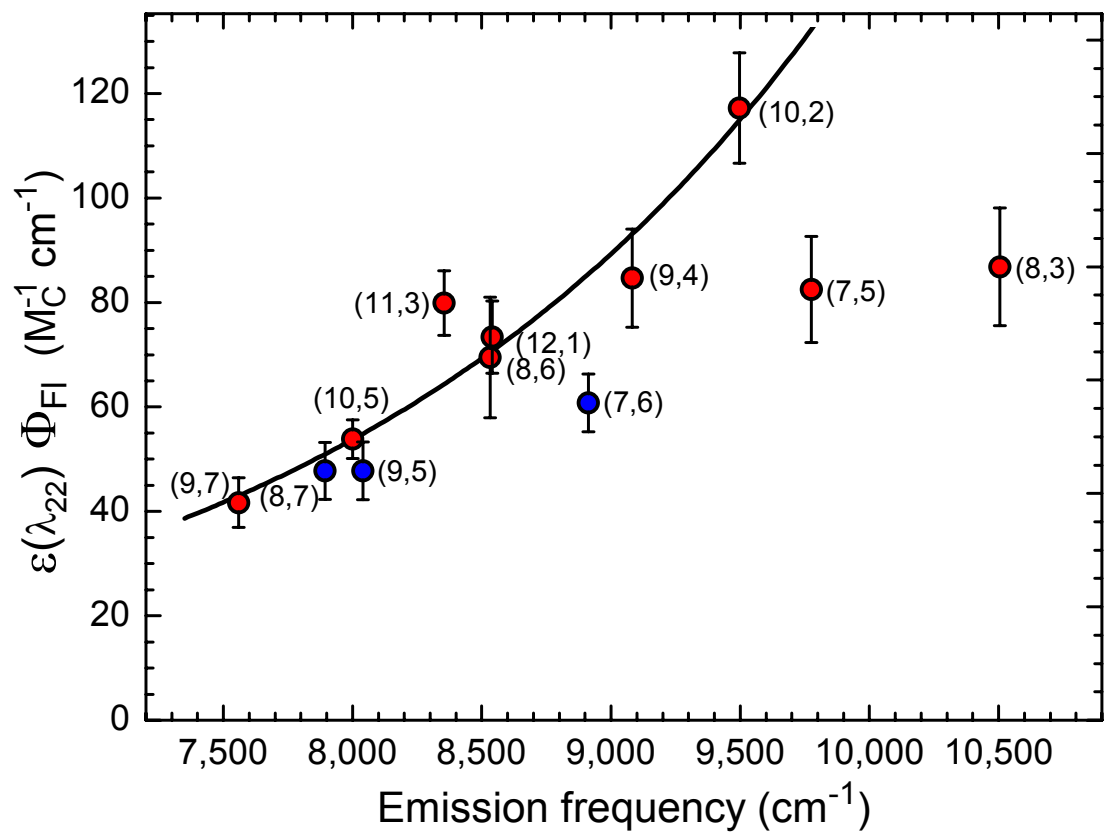

Figure 2. Dependence of fluorescence action cross-section on emission frequency, as measured for 12 ($n,m$) species. The ordinate is expressed in units of molar absorptivity (base 10) referred to moles of carbon. Blue symbols indicate mod($n-m$,3)=1; red symbols indicate mod($n-m$,3)=2. Error bars show the standard error of the mean for each species. As described in the text, the solid curve is a model fit to all points except (7,6), (7,5), and (8,3).

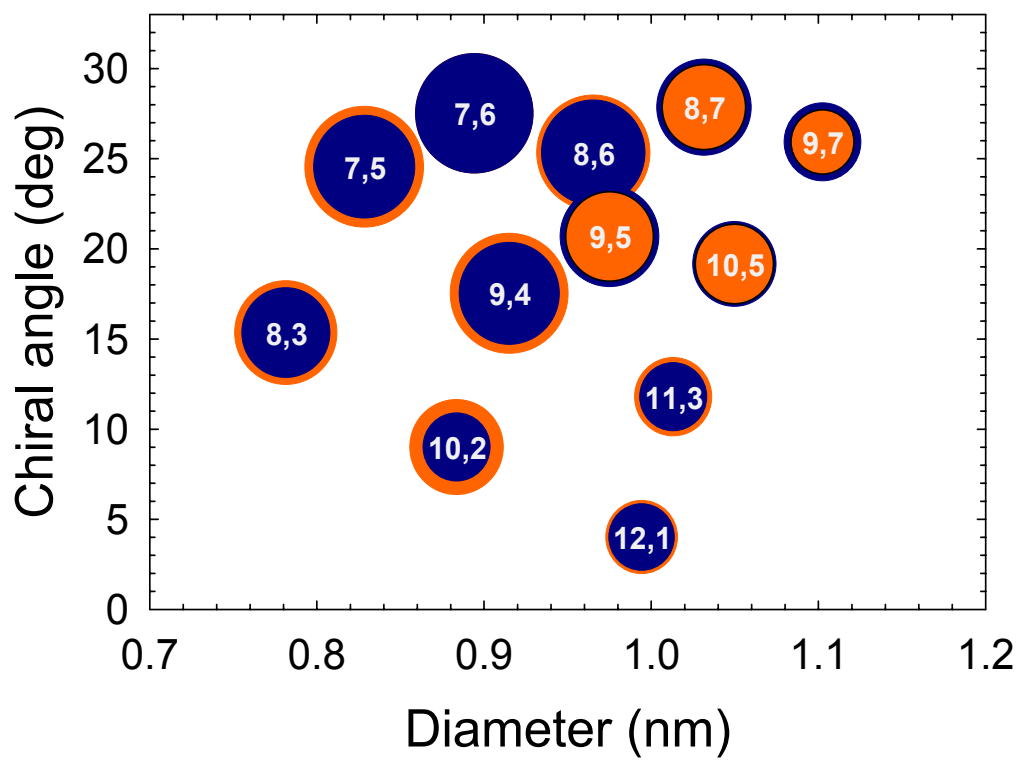

Figure 3. Relative (*n*,*m*)-resolved fluorimetric distributions from a sample of HiPco SWNTs (Rice HPR-89) in aqueous SDBS suspension. The area of each (*n*,*m*)-labeled circle is proportional to the corresponding signal magnitude. Orange (light) and blue (dark) circles represent signals before and after correction by the action cross-sections listed in Table 1. The two distributions are normalized to match the (7,6) signals.

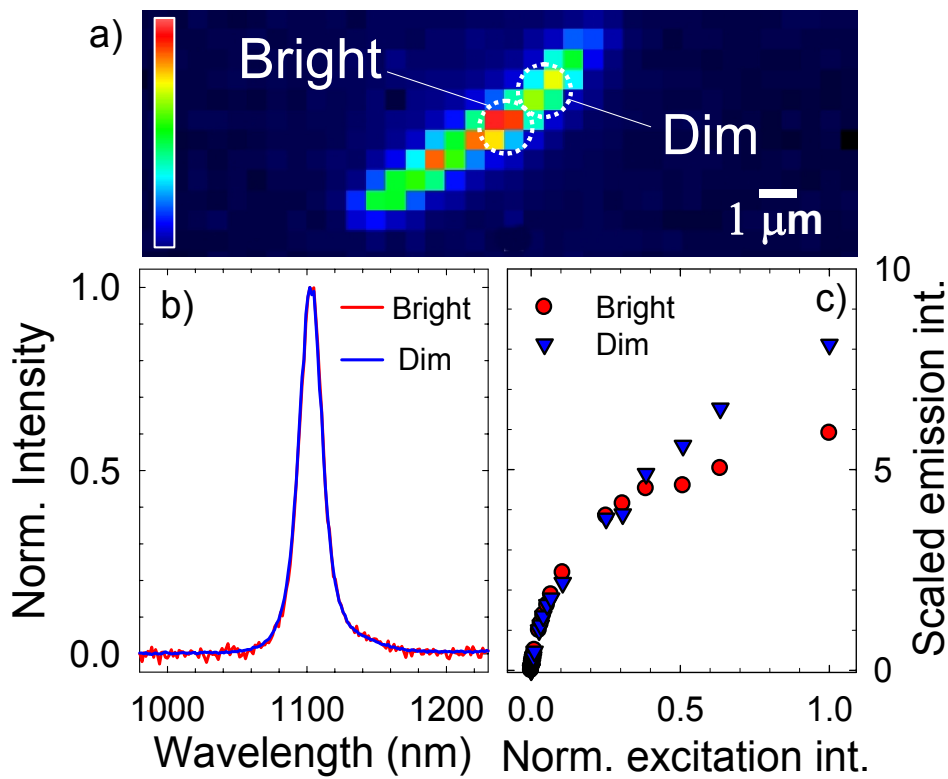

Figure 4. Comparison of fluorescence from adjacent bright and dim segments of the same SWNT. (a) Fluorescence image of the (7,6) nanotube at 90× magnification with the selected segments circled. (b) Overlaid normalized emission spectra recorded from these two segments. (c) Dependence of emission intensity on normalized excitation intensity for the two segments, after y-axis scaling to match the low excitation intensity regions. The maximum excitation intensity was $1.2 \pm 0.2$ kW/cm$^2$.

Table of Contents Figure

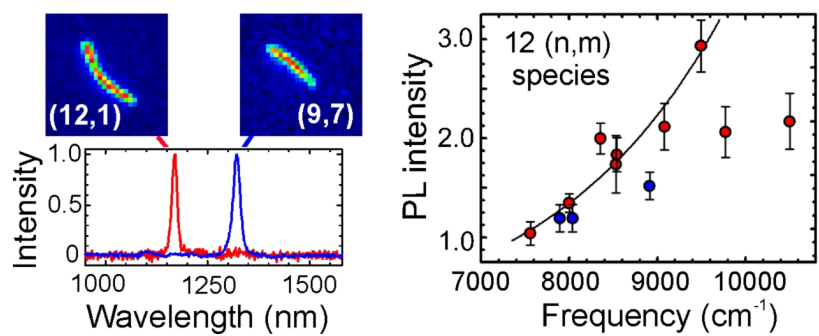